# Macroengineering in the Galactic Context: A New Agenda for Astrobiology

**Milan M. Ćirković**

*Astronomical Observatory of Belgrade*
*Volgina 7, 11160 Belgrade, Serbia and Montenegro*
*e-mail:* mcirkovic@aob.bg.ac.yu

Abstract: We consider the problem of detectability of macro-engineering projects over interstellar distances, in the context of Search for ExtraTerrestrial Intelligence (SETI). Freeman J. Dyson and his imaginative precursors, like Konstantin Tsiolkovsky, Olaf Stapledon or John B. S. Haldane, suggested macro-engineering projects as focal points in the context of extrapolations about the future of humanity and, by analogy, other intelligent species in the Milky Way. We emphasize that the search for signposts of extraterrestrial macro-engineering projects is not an optional pursuit within the family of ongoing and planned SETI projects; inter alia, the failure of the orthodox SETI thus far clearly indicates this. Instead, this approach (for which we suggest a name of "Dysonian") should be the front-line and mainstay of any cogent SETI strategy in future, being significantly more promising than searches for directed, intentional radio or microwave emissions. This is in accord with our improved astrophysical understanding of the structure and evolution of the Galactic Habitable Zone, as well as with the recent wake-up call of Steven J. Dick to investigate consequences of postbiological evolution for astrobiology in general and SETI programs in particular. The benefits this multidisciplinary approach may bear for macro-engineers are also briefly highlighted.

**Introduction**
Living beings change their environments and are changed by their environments in turn. This truism has become especially pertinent within the framework of astrobiology. Even before the onset of the explosive development of this field we are witnesses of, the fact that even simple lifeforms can influence its physical and chemical environment on the planetary scale has been widely known. The stock example is the one of the Earth's atmosphere, which is markedly out of chemical equilibrium due to the presence of the biosphere, and has been so for billions of years. As the author of the Gaia hypothesis, James Lovelock wrote: "Almost everything about its composition seems to violate the laws of chemistry... The air we breathe... can only be an artifact maintained in a steady state far from chemical equilibrium by biological properties." (Lovelock 1988) Recently publicized projects dealing with detection of exoplanetary biospheres all rely on this simple fact. In somewhat different light, we are all sadly aware of the impact of human activities on the biosphere, Earth's climate and Earth's circumplanetary space.
Even more ambitious in this regard are projects falling in the wide spectrum of views and approaches characterizing the new multidisciplinary topic of macro-engineering. All these dynamic approaches are worthy successors to the ultimate goal set in Francis Bacon's posthumous New Atlantis (1626): "The end of our foundation is the knowledge of causes, and secret motions of things; and the enlarging of the bounds of human empire, to the effecting of all things possible." Here we shall attempt to offer a new twist on the same old theme, a twist that is likely to open multiple new and unexpected vistas of research. In order to expose this novel perspective, we need to take a brief detour through a seemingly unrelated, but in fact quite relevant and vigorously expanding discipline.

We are lucky enough to live in an epoch of great progress in the nascent field of astrobiology, which deals with three canonical questions: How does life begin and develop? Does life exist elsewhere in the universe? What is the future of life and intelligence on Earth and in space? A host of discoveries have been made during the last decade or so, the most important certainly being a large number of extrasolar planets; the existence of many extremophile organisms at the deep ocean hydrothermal vents, possibly vindicating the "deep hot biosphere" of Thomas Gold (1998); the discovery of subsurface water on Mars and the huge ocean on Europa, and possibly also Ganymede and Callisto; the unequivocal discovery of amino-acids and other complex organic compounds in meteorites; modelling organic chemistry in Martian and Titan's atmosphere; the quantitative treatment of the Galactic Habitable Zone; the development of a new generation of panspermia theories, spurred by experimental verification that even terrestrial microorganisms survive conditions of an asteroidal or a cometary impact; progress in philosophy and methodology, etc. (for recent beautiful reviews see Des Marais and Walter 1999; Darling 2001; Grinspoon 2003). All this deserves the label of a true astrobiological revolution.

Perhaps the most fascinating field in the multidisciplinary astrobiological spectrum is the Search for ExtraTerrestrial Intelligence (henceforth SETI). At the beginning of XXI century it remains the oldest and perhaps the most intriguing scientific problem. Starting with the pioneering work of Frank Drake, Carl Sagan, Iosif Shklovsky, and others (e.g., Drake, 1965), as well as the historical OZMA project, SETI studies have had their ebb and flow of tides over the last four decades (Dick, 1996). During that time a set of ideas which can be characterized as "orthodox" SETI has emerged. In a simplified form, it can be summarized as follows. Life is common in the universe. Emergence of intelligence and technology is, if not necessary, then at least typical outcome of biological evolution throughout the Milky Way. A sizeable fraction of technological species are interested in communication with other intelligent creatures. It makes sense to listen for intentional radio or optical messages from out there and to transmit messages in return. It makes no sense to travel across interstellar distances or to expect such interstellar visitors. What we can hope to achieve is slow and benign exchange of messages, the greatest beneficiaries in such exchange being the youngest newcomers to the "Galactic Club" (Bracewell, 1975), such as humans. Basic tenets of this view have crystallized by mid-1970s, decades before the astrobiological revolution.

Times are changing. In addition to the astrobiological revolution itself, some of the important recent developments of relevance to the SETI endeavour (in the widest sense) are:

The rise of digital perspective in various fields, starting with fundamental physics and computer science (e.g., Chaitin, 1987; Toffoli, 1998; Chaisson, 2001; Fredkin 2003), to biological and social sciences (e.g., Kauffman, 1995; Maynard Smith and Szathmary, 1997; Adami et al., 2000; Carroll, 2001; Adams, 2003). In particular, this includes understanding of the crucial importance of information dynamics, open systems, complexity, substrate-independent dynamical laws, and interrelated evolutionary pathways.

Closely related, tremendous efforts in the fields of computer science and neurosciences, invested toward achieving of artificial intelligence (AI), which would offer completely new perspectives on the nature of intelligence itself (Henry, 2005), as well as the possible future evolutionary trajectory of humanity and, by analogy, other intelligent communities in the Galaxy. Coupled with the digital perspective in biological science, this raises the all-important (from the SETI point of view) issues of whether we should search for biological or postbiological intelligence; this is particularly forcefully put forward in a recent important paper by the distinguished historian and philosopher of science Steven J. Dick (2003). Information theory also recently highlighted all the difficulties (Lachmann, et al. 2004) and inefficiencies (Rose and Wright, 2004) inherent in the attempts to communicate by radio signals over interstellar distances.

The advent of physical eschatology, a nascent astrophysical discipline dealing with the future of astronomical objects, including the universe itself. The groundworks were laid by Freeman J. Dyson a quarter century ago (Dyson, 1979), but the explosion of interest occured only in the last decade or so; for reviews see Adams and Laughlin (1997), Ćirković (2003).

Related to research in physical eschatology, though on more modest spatio-temporal scales are those aspects of future studies dealing with the future of humanity, and, in particular, the rise of new vigorous intellectual movements, which can be (in a revived term of Sir Julian Huxley; see Huxley, 1957) unified under the banner of transhumanism (Moravec, 1988; Kurzweil, 2000; Wright, 2000; Bostrom, 2005). These may signify an open endorsement of the postbiological paradigm, to which we shall return later.

Improved epistemological and methodological understanding of our properties as intelligent observers, as well as physical, chemical and other pre-conditions necessary for the existence of such observers (Carter, 1983; Livio, 1999). The latter are topics of so-called anthropic reasoning, the subject of much debate and controversy in cosmology, fundamental physics, and philosophy of science (Barrow and Tipler, 1986). In recent years it became clear that anthropic principle(s) can be most fruitfully construed as observation selection effects (Bostrom, 2002). This is a straightforward continuation of the Copernican Revolution, which emphasizes a non-special character of our cosmic habitat, and which has so immensely contributed to our scientific understanding.

The rise of neocatastrophism in Earth and planetary sciences, offering not only wealth of explanatory models and hypotheses for puzzling empirical facts (e.g., Jablonski, 1986; Raup, 1991, 1994; Courtillot, 1999; Benton, 2003), but a new philosophical perspective as well (Gould, 1985; Huggett, 1997).

Emergence of serious macro-engineering studies on various scales of space, time, energy and sophistication, best attested by the other contributions in this volume. (Let us note in passing that macro-engineering, as by definition the intentional change of physical environment on ever larger scales, is antithetical to the main goal of physical eschatology, which tries to track future evolution of our cosmic environment with as little complication due to the intentional influences. The epistemological and methodological relationship between the two has not been sufficiently investigated in the literature so far, although there have been several interesting studies at the crossroads; e.g., Garriga et al., 2000; Ćirković, 2004.) In particular, the aspects of macro-engineering pertaining to our astronomical surroundings (e.g., Korycansky, et al. 2001; McInnes, 2002) are of relevance for our search for intelligence elsewhere.

If we now wish to ask how the actual SETI research has responded to these exciting and stimulative developments of obvious relevance to its subject matter, we are bound for severe disappointment. With honorable exceptions, the mainstream SETI (as exemplified, for instance, by activities of NASA, the SETI Institute, or the Planetary Society; e.g., Shostak 1993; Pierson 1995; McDonough 1995) has mostly ignored all of the developments listed above. SETI today is mostly in the same shape and with the same set of philosophical, methodological and technological guidelines, as it was in the time of its pioneers (Frank Drake, Carl Sagan, Giuseppe Cocconi, Philip Morrison, Iosif Shklovsky, Ronald Bracewell, Bernard Oliver, Michael Papagiannis, Nikolai Kardashev) in 1960s and 1970s. In contrast which can hardly be sharper, our views of astrophysics, planetary sciences, evolutionary biology and, especially, computer science—arguably the four key scientific pillars of SETI—changed revolutionarily, to put it mildly, since that epoch. It is both sad and ironic that the field which was once correctly identified as a paragon of originality, boldness, and vigour, has not lived to these ideals during the last 30 years or so; instead, it was gradually subverted by conservative views. Only by this odd conservativeness can be explained that, for instance, a major SETI review article at the beginning of the new millenium in by far the most authoritative publication in astronomical sciences (Tarter, 2001) can be written without even a single mention (in its 38 pages) of such crucial memes as AI, anthropic reasoning, von Neumann probes, neodarwinism, or macro-engineering. Claims that (orthodox) SETI is a strictly empirical field, should be evaluated at face value; translated into the language of modern epistemology, this would serve the counterproductive role of sorting SETI with pseudo-sciences, since amassing "empirical work" while having no theoretical basis (or hiding it) is a hallmark of pseudo-science. (Compare with cases of parapsychology, dowsing, craniometry or various nutritionisms, including the ideas of undoubted geniuses, like Linus Pauling.) The present study is an attempt to break this mold and point to serious modern alternatives to the old-fashion SETI philosophy. This is not a luxury, but a necessity in the situation when epochal changes in our views of virtually everything in science and technology are not, unfortunately, accompanied by an appropriate shift in SETI attitudes.

This sort of isolationism is never justified, not only in science, but in wider arena of intellectual life; however, as history teaches us, it could be partly mitigated by inherent short-term successes of the isolated discipline. This is certainly not the case with SETI studies. In almost four decades of SETI projects there have been no results, in spite of the prevailing "contact optimism" of 1960s and 1970s, motivated largely by uncritical acceptance of the Drake equation (Drake, 1965). Conventional estimates of that period spoke about $10^6 – 10^9$ advanced societies in the Milky Way forming the "Galactic Club" or a similar anthropocentrically conceived association; for a prototype optimistic—or naive—view of that epoch, exactly 30 years old, see Bracewell (1975). Early SETI literature abounds in such misplaced enthusiasm. Today, even "contact-optimists" have abandoned fanciful numbers, and settled on a view that advanced extraterrestrial societies are much rarer than previously thought. One of the important factors in this downsizing of SETI expectations has been demonstrations by "contact pessimists", especially Michael Hart and Frank Tipler, that the colonization—or at least visit—of all stellar systems in the Milky Way by means of self-reproducing von Neumann probes is feasible within a minuscule fraction of the Galactic age (Hart, 1975; Jones, 1976; Tipler, 1980, 1981). In this light, Fermi's legendary question: Where are they? becomes disturbingly pertinent (Brin, 1983; Webb, 2002). In addition, Carter (1983) suggested an independent and powerful anthropic argument for the uniqueness of intelligent life on Earth in the Galactic context, and biologists such as Simpson (1964), Mayr (1993), or Conway Morris (2003) put forward biological arguments to the effect that complex lifeforms are quite rare in the Galaxy. These views have sometimes been subsumed within the "rare Earth" hypothesis (Ward and Brownlee, 2000). It is generally recognized in both research and the popular science circles that the "contact pessimists" have a strong position.

While fully recognizing that patience is a necessary element in any search, cosmic or else, we still wish to argue that the conventional SETI (Tarter, 2001; Duric and Field, 2003, and references therein), as exemplified by the historical OZMA Project, as well its subsequent and current counterparts (META, ARGUS, Phoenix, SERENDIP/Southern SERENDIP, etc.), notably those conveyed by NASA and the SETI Institute, is fundamentally limited and unlikely to succeed. This is emphatically not due to the real lack of targets, us being alone in the Galaxy, as contact-pessimists in the mold of Tipler or Mayr have argued. Quite contrary, it is due to real physical, engineering, and epistemological reasons undermining the conventional SETI philosophy. In a sense the problem has nothing to do with the universe itself, and everything to do with our ignorance and prejudices.

In particular, we wish to argue that the reality is not limited to these two extremes: (I) naive optimism of the SETI establishment inherited from the "founding fathers", and (II) blatant pessimism of the detractors, often supported by extra-scientific motives. (Examples of the latter are worrying about the magnitude of the U.S. federal debt in Mayr (1993), or various (quasi)religious elements present, e.g., in Conway Morris (2003) or in Tipler's numerous writings. Darling (2001) amusingly discusses the relationship of the "rare Earth" hypothesis of Ward and Brownlee, a mainstay of SETI skepticism today, with creationism.) As usual, the reality is much more complex, and taking into account the developments listed above will give rise to a set of more complex "middle-ground" solutions to the problem of absence (or otherwise) of extraterrestrial intelligence. Such hypotheses are, typically, not open to falsification with the standard SETI procedures (i.e., listening to intentional radio- or optical messages), but are, in general, falsifiable by a different set of SETI methods and procedures. One such "middle-ground" solution has been proposed by the present author elsewhere (Ćirković and Bradbury, 2005), but it is far from being unique in this respect; a large class of "catastrophic" solutions to Fermi's paradox belong to this category (Clarke, 1981; Annis, 1999a), as well as those based on the long-term evolutionary processes (Schroeder, 2002; Ćirković, 2005). This also applies to the ingenious idea that advanced civilizations will transfer their cognition into their environment (Karl Schroeder, private communication), following recent studies on the distributed natural cognition (e.g., Hutchins, 1996). In these, as in other suggested lines of postbiological evolutionary development of advanced civilizations, the approaches favored by the ongoing SETI projects will be fundamentally misguided, i.e., advanced societies remain undetectable by such approaches. All of these ideas require a sort of "breaking the mold" of conservative mainstream thinking.

Such "breaking the mold" must not be understood as leading inexorably to increased SETI optimism. A very good counterexample is the work of Raup (1992) on non-conscious SETI sources, which are likely to cause confusion in the practical SETI work. Although we hereby argue that on the balance, changed perspectives increase chances of actual SETI success, this is rather accidental, certainly not a necessary consequence. On the contrary, many skeptical arguments can and must be incorporated in the emerging "new" SETI paradigm.

The crucial ingredient here is exactly our development #7 above, namely the increased awareness of the potentials inherent in macro-engineering. The central point was clearly explicated long ago by Freeman J. Dyson, to whose ideas we shall repeatedly return (Dyson, 1966, p. 642):

When one discusses engineering projects on the grand scale, one can either think of what we, the human species, may do here in the future, or one can think of what extraterrestrial species, if they exist, may have already done elsewhere. To think about a grandiose future for the human species ("la nostalgie du future") is to pursue idle dreams, or science fiction. But to think in a disciplined way about what we may now be able to observe astronomically, if it should happen to be the case that technologically advanced species exist in our corner of the universe, is a serious and legitimate part of science... In this way I am able to transpose the dreams of a frustrated engineer into a framework of respectable astronomy.

While we may take a votum separatum on some of Dyson's views on future studies (written almost 40 years ago), the major point of this passage seems almost self-evident: one of the best ways to ascertain limits of the technically possible is to discard anthropocentrism and sample a larger volume of space and time, reasonably expecting that somewhere and somewhen it has been achieved. In the rest of this Chapter, we shall consider epistemological and methodological impact of macro-engineering (sometimes called astro-engineering in this context; we shall use the terms synonymously) on our SETI projects and policies. As we hope to show, this contact between macro-engineering and astrobiology can be encouraging and productive on both sides.

Dysonian Approach to SETI

By their fruits ye shall know them. The Biblical proverb neatly encapsulates the proposed unconventional approach to SETI, in which the focus would be a search for manifestations and macro-engineering artifacts, instead of intentional messages. Even more, the metaphor seems particularly apt, since it warns about messages (and efforts to hear them) being actively misleading in search for the truth.

It was along these lines that, in 1960, Freeman J. Dyson suggested that the very existence of what we can term advanced technological civilizations (henceforth ATCs) should provide us with means of detecting them (Dyson, 1960). As is well-known, starting from the Malthusian assumptions and the well-recorded increase in power consumption with the development of technological civilization here on Earth, Dyson concluded that a truly developed society will soon face the limits regarding both living space and available energy if constrained only to planetary surfaces. On the contrary, the only way to optimize resources would be a construction of a Dyson shell, capturing all energy from the domicile star. This particular paper, hardly longer than a page, has not only motivated manifold subsequent visions and studies in the field of astro-engineering and is likely to continue to do so for a long time to come; we argue that it set groundwork for a different sort of SETI from the one conducted since the OZMA project in mid-1960s. That is because Dyson suggested that infrared signatures of a Dyson shell will be detectable from large distances, and will present a confirmation of the existence of ATCs. This view has been subsequently elaborated in Dyson (1966), and in the often-neglected study of Sagan and Walker (1966). Some of the very best elaborations of the Dysonian ideas have been published in the science fictional context by Stanislaw Lem (1984, 1987; for a discursive form, see Lem 1977). For further fruitful work along Dyson's lines, see Chapter 14 of the present book.

(It is neither necessary nor desirable for our further considerations to make the notion of ATCs more precise. The diversity of postbiological evolution is likely to at least match, and probably dwarf, the diversity of its biological precedent. It is one particular feature—information processing—we assume common for the "mainstream" ATCs, in accordance with the postbiological paradigm and the Intelligence Principle of Dick (2003). Thus, whether real ATCs can most adequately be described as "being computers" or "having computers" is not of key importance for our analysis; we just suppose that in either case the desire for optimization of computations will be one of important (if not the most important) desires of such advanced entities. It is already clear, from the obviously short and limited human experience in astronautics, that the postbiological evolution offers significant advantages in this respect (Parkinson, 2005). However, an operational definition of ATC is clear: ATC is the community capable of grand feats of astro-engineering. This is another important link between the fields of astrobiology and macro-engineering, which certainly requires further elaboration elsewhere.)

It is thus multiply justified to call this other, unconventional paradigm the Dysonian approach to the problem of extraterrestrial intelligence. Some of its elements were in place, to be sure, much before Dyson and his 1960 paper. Notable sources for a future historian of ideas would be far-reaching speculations of several brilliant minds in the first few decades of the XX century: British authors and philosophers William Olaf Stapledon (1886-1950) and Herbert George Wells (1866-1946), as well as the famous biologist and polymath John B. S. Haldane (1892-1964); Russian engineer and astronautics pioneer Konstantin E. Tsiolkovsky (1857-1935); Serbian-American inventor and electrical engineer Nikola Tesla (1856-1943); as well as some others. It is impossible to give justice to the magnitude and boldness of thought of these great minds in this limited space. Let us just mention in passing that Stapledon, Wells and Haldane mused presciently upon great powers which will be available to future man in shaping both himself and his natural environment; in addition, they emphasized the need for synergistic development of both natural and social sciences as well as technology to achieve such powers, without succumbing to a catastrophe on the way. Tsiolkovsky was perhaps the first macro-engineer in the modern sense, envisioning huge space colonies and habitats (the idea later developed by Gerard O'Neill and others) serving as springboards for human space expansion and colonization; he was also a precursor to the Fermi paradox, this most perplexing SETI-related puzzle. While Tesla can claim credit for being the founder of orthodox SETI by first emphasizing the role of radio-waves and even doing some practical attempts to detect alien radio signals (Dick, 1996), he was also the first to perceive the link between the stage of development of a planetary society and a specific physical quantity – namely expended power. In an artistic manner, the same inspiration of macro-engineering in the cosmical context motivated some of the best graphics of one of the most prominent artist of the XX century, Maurice C. Escher (see Figures 13-1, 2).

But it was Dyson, "the eclectic physicist and the frustrated engineer" by his own amusing description in a later work (Dyson, 1966), who provided the key insight which, unfortunately, almost half a century later has not been sufficiently understood in all its ramifications and consequences. It is important to emphasize here that the Dysonian approach to SETI is not limited to search for Dyson shells, but to general class of artifacts, manifestations and traces of the existence of ATCs. Some other astro-engineering feats belonging to this, potentially detectable, category are:

Supramundane planets, shell worlds, orbital rings, and similar circumplanetary constructions (Birch 1983, 1991; Roy, et al. 2004).
Large-scale antimatter-burning vehicles or industrial plants (Harris, 1986, 2002).
Large artificial objects (Tsiolkovsky-O'Neill habitats, for instance) in transiting orbits, detectable through extrasolar planet searches (Arnold, 2005).

Even the search for extraterrestrial artifacts in the Solar System belongs to this category. This approach, however, is still strongly in the minority in SETI circles, in contrast to its many advantages, some of which we shall briefly review.

The very obvious thing is that such an approach does not prejudice properties of target societies in the way the orthodox view does. It is self-evident that willingness of both parties is the necessary (but far from sufficient) condition for a successful communication on any scale, from human everyday life up. However, in contrast to people we encounter every day in our lives, we have not the vaguest idea of whether such willingness exists on the level of interstellar communication. It is indicative that a large portion of the early SETI literature, especially writings of the "founding fathers" consists of largely emotional attempts to make the assumption of willingness (and, indirectly, benevolence) of SETI target societies plausible (e.g., Bracewell, 1975); this is read more like wishful thinking than any real argument (Gould, 1987). To cite Dyson (1966) again: "[M]y point of view is rather different, since I do not wish to presume any spirit of benevolence or community of interest among alien societies." This, of course, does not mean that the opposite assumption (of malevolence) should be applied. Simply, such prejudicating in the nebulous realm of alien sociology is unnecessary in the Dysonian framework; with fewer assumptions it is easier to pass Occam's razor.

The most important advantage of the Dysonian approach concerns the spatial and temporal frames within which practical SETI is conducted. Even proponents of the SETI orthodoxy admit that the "window of opportunity" for radio- or optical laser communication is quite short; in a cogent and well-written summary of the orthodox position, Duric and Field (2003) admit:

Under the optimistic scenario, there may be as many as 106 technologically advanced civilizations in the Galaxy. However, these societies are at various stages of development. The probability that two extraterrestrial societies are at the same stage of evolution, to say within a million years, is very small.

This is obviously of crucial importance if our goal is bidirectional, intentional communication between us and an extraterrestrial civilization. It is exactly to this situation that the arguments of SETI skeptics most forcefully apply; thus, it is important to realize that what is often vaguely referred to as "anti-SETI" arguments are, indeed, only arguments against SETI orthodoxy. However, the same "window of opportunity" is increased by many orders of magnitude, or even vanishes entirely, when we search for macro-engineering artifacts instead of intentional messages. This reasoning was clear to thinkers of epochs long past. In a beautiful passage in Book V of the famous poem De Rerum Natura [On the Nature of Things], Roman poet and late-Epicurean philosopher Lucretius wrote the following intriguing verses (in translation of William E. Leonard, available via WWW Project Gutenberg; Lucretius 1997):

If there had been no origin-in-birth
Of lands and sky, and they had ever been
The everlasting, why, ere Theban war
And obsequies of Troy, have other bards
Not also chanted other high affairs?
Whither have sunk so oft so many deeds
Of heroes? Why do those deeds live no more,
Ingrafted in eternal monuments
Of glory? Verily, I guess, because

The Sun is new, and of a recent date
The nature of our universe, and had
Not long ago its own exordium.

Neglecting here its cosmological context of arguing for a finite past age of the universe, this passage indicates an oft-neglected aspect of Fermi's paradox—it is not enough to somehow remove all ATCs from our past light cone, but we need to erase their more durable and potentially detectable achievements as well, in order to reproduce empirical "Great Silence" (Brin, 1983). On Earth, the very existence of the fascinating discipline of archaeology tells us that cultures (and even individual memes) produce records significantly more durable than themselves. It is only to be expected that such trend will continue to hold even more forcefully for higher levels of complexity and more advanced cultures. There are even some factors related to the properties of our cosmic environment that enhance this trend; notably, it has already been repeatedly suggested that the traces of any hypothetical extraterrestrial visitations in the past of the Solar System would be easier to locate on the Moon than on Earth, due to the vastly supressed erosion there (e.g., McKay, 1996).

Figure 1. Astro-engineering as inspiration in arts: M. C. Escher's Concentric Rinds (1953) can be regarded as Dyson shell-like structures (in construction?).

As an example, let us for the sake of discussion allow that a significant fraction of advanced technological civilizations evolves toward the Kardashev (1964) Type II, i.e. a community completely managing the energy output of its parent star; for the information-processing need of advanced communities, see Ćirković and Radujkov (2001). The straightforward way of achieving this is the construction of a Dyson shell. Once constructed, such an example of astroengineering, will be quite durable due to the properties of the interplanetary and interstellar space itself; like the Pyramids of Egypt, a Dyson shell is likely to outlive its creators for a vast period of time (on the physical eschatological scales), thus being an advanced analog of Lucretius' "eternal monuments". Some very preliminary searches (see §3 below) show the absence of such artifacts in the Galactic vicinity of the Sun.
Similar reasoning can be applied to the volume of space sampled by active searches. According to recent important studies by Lineweaver (2001) and Lineweaver, et al. (2004), Earth-like planets around other stars in the Galactic Habitable Zone (Gonzalez et al., 2001) are, on average, 1.8 ± 0.9 Gyr older than our planet. These calculations are based on chemical enrichment as the basic precondition for the existence of terrestrial planets, as well as on the rate of destructive processes (like supernovae). Applying the Copernican assumption naively, we would expect that correspondingly complex life forms on those others to be on the average 1.8 Gyr older. Intelligent societies, therefore, should also be older than ours by the same amount. In fact, the situation is even worse, since this is just the average value, and it is reasonable to assume that there will be, somewhere in the Galaxy, an inhabitable planet (say) 3 Gyr older than Earth. Since the set of intelligent societies is likely to be dominated by a small number of oldest and most advanced members (for an ingenious discussion in somewhat different context, see Olum, 2004), we are likely to encounter a civilization actually more ancient than 1.8 Gyr (and probably significantly more). It seems preposterous even to contemplate any possibility of communication between us and Gyr-older supercivilizations. Remember that 1 Gyr ago the appearance of even the simplest animals on Earth lay in the distant future. Some of the SETI pioneers have been very well aware of this on the qualitative level and warned about it (notably Sagan, 1975); these cautious voices have been consistently downplayed by the SETI community. All in all, we conclude that the conventional radio SETI assuming beamed broadcasts from a nearby Sun-like star (e.g., Turnbull and Tarter, 2003) is ill-founded and has almost no chance of success on the present hypothesis. Given the likely distances of ATCs that began their technological ascent tens of millions to billions of years ago, they are not likely to know of our development. While their astronomical capabilities probably allow them to observe the Solar System, they are looking at it before civilization developed. It is doubtful, to say at least, that they would waste resources sending messages to planetary systems possessing life, but quite uncertain (in light of the biological contingency) to develop a technological civilization. Dolphins and whales are quite intelligent and possibly even human-level conscious (e.g., Browne, 2004), but they do not have the ability to detect signals from ATCs, and it is unlikely, to say at least, that they will ever evolve such a capacity. By a mirror-image of such position, unless one has concrete evidence of an ATC at a given locale it would be wasteful to direct SETI resources towards them. Although this conclusion can offer a rationale to some of the SETI skeptics, it is based on the entirely different overall astrobiological picture and has different practical consequences.

Furthermore, locations of the original home and the bulk of technology of an ATC can be decoupled even at smaller spatial scales. It has already been repeatedly suggested that our descendants, in particular if they cease to be organic-based, may prefer low-temperature, volatile-rich outer reaches of the Solar System (if these ecological niches are not already filled; see Dyson, 2003). Thus, they could create what could be dubbed a "circumstellar technological zone" as different and complementary to the famous "circumstellar habitable zone" in which life is, according to most contemporary astrobiological views, bound to emerge. We propose to generalize this concept to the Galaxy (and other spiral galaxies) in complete analogy to the Galactic habitable zone.

In addition, the Dysonian approach allows us to re-assess extragalactic SETI, in the sense precluded by the orthodox paradigm. Extragalactic SETI has not been considered very seriously so far (for notable exceptions see Wesson 1990; Annis 1999b). The reason is, perhaps, the same old comforting prejudice that we should expect specific (and most conveniently radio) signals. Since these are not likely forthcoming over intergalactic distances and the two-way communication desired by orthodox SETI pioneers is here entirely senseless, there is no point in even thinking seriously about extragalactic SETI. Such view is fallacious: when we remove the cozy assumption of specific SETI signals (together with the second-order assumption of their radio nature), it collapses. On the contrary, extragalactic SETI would enable us to probe enormously larger part of physical space as well as the space of possible evolutionary histories of ATCs. (Of course, part of what we get ensemble-wise we lose time- and resolution-wise.) In fact, the definition of Kardashev's Type III civilization (i.e., those managing energy resources of its entire home galaxy; Kardashev, 1964) should prompt us to consider it more carefully, at least for a sample of nearby galaxies, visible at epochs significantly closer to us than the 1.8 Gyr difference between the average of Lineweaver (2001) and the age of Earth which is about 4.56 Gyr. In fact, it could be argued (although it is beyond the scope of the present study) that the null result of extragalactic SETI observations so far represents a strong argument against the viability of Kardashev's Type III civilizations. While it remains a possibility, in the formal sense of being in agreement with the known laws of physics, it seems that the type of pan-galactic civilization as envisaged by Kardashev and other early SETI pioneers is either much more difficult (suggesting that the sample of ~104 normal spiral galaxies close enough and observed in high enough detail is simply too small to detect even a single Type III civilization), or simply not worth striving to establish.

**A prolegomena for "new" seti**

The list of both theoretical and observational SETI studies performed so far along the Dysonian guidelines is rather short; most of it is the following:

Searches for artificial objects near Earth and anomalous spectral lines in stellar spectra performed by Robert A. Freitas, Jr. and Francisco Valdes in 1980s (Freitas and Valdes 1980; Valdes and Freitas 1983, 1986).

Japanese program of searches for Dyson shells around nearby stars (Jugaku, et al. 1995; Jugaku and Nishimura, 2003).

Proposals for observational or archival searches for Dyson shells or related astro-engineering projects (Slysh, 1985; Tilgner and Heinrichsen, 1998; Timofeev, et al. 2000).

The detailed study of Sandberg (2000), fruitfully linking information-processing to macro-engineering.

Investigation of gamma-ray signatures of antimatter burning by ATCs (Harris, 1986, 2002).

A recent proposal for searching for transits of artificial objects across the observed stars (Arnold, 2005).

The analysis of archival extragalactic data by Annis (1999b) suggesting the absence of star-powered Kardashev Type III civilizations among nearby galaxies.

To this disturbingly short list one may add several important theoretical studies showing either general feasibility of the astro-engineering feats detectable from afar (e.g., Suffern, 1977; Badescu, 1995; Badescu and Cathcart, 2000), or the necessity of taking a non-standard approach (e.g., Russell, 1983; Raup, 1992); the latter have been, significantly enough, often written by biologists interested in SETI, and have not been given due credit and attention in the orthodox SETI circles.

Figure 2. Another Escherian vision of a macro-project: woodcut Tetrahedral Planetoid (1954) reminiscent of much later O'Neill's orbital habitats. M. C. Escher was an enthusiastic amateur astronomer, who meticulously observed double stars and other celestial phenomena (cf. Locher et al., 1992).

The proposed re-orientation of SETI projects is in a very deep sense independent of the favorite model for solution of puzzles related to the extraterrestrial intelligence, most notably Fermi's paradox. Although there may indeed be more than fifty solutions to Fermi's paradox (Webb, 2002), essentially all major solutions are compatible with, or indeed suggestive of the Dysonian approach.

Of course, even those projects or proposals put forward so far are limited in the sense of being often too conservative with respect to the full range of parameters. For instance, the controlling parameter for detection of a Dyson shell is, ultimately, its temperature (the differences in the intrinsic stellar output can be neglected in the first approximation). The searches thus far (including the mentioned studies of Jugaku et al.) relied on the original Dyson's proposal that the shell would be the size of Earth's orbit around the Sun, and that its working temperature would, thus, be close to the temperature of a solid body at 1 AU from a G2 dwarf. However, from a postbiological perspective, this looks to be quite wasteful, since computers operating at room temperature (or somewhat lower) are limited by a higher kT ln 2 Brillouin limit, compared to those in contact with heat reservoir on lower temperature T (cf. Brillouin, 1962). Although it is not realistic to expect that efficiency can be increased by cooling to the cosmological limit of 3 K in the realistic model of the Galaxy (Ćirković and Bradbury, 2005), still it is considerable difference in practical observational terms whether one expects a Dyson shell to be close to a blackbody at 50 K, as contrasted to a blackbody at 300 K. This lowering of the external shell temperature is also in agreement with the study of Badescu and Cathcart (2000) on the efficiency of extracting work from the stellar radiation energy. In this sense, the "true" Dysonian approach needs to be even more radical than the intuitions of Dyson himself.

This is linked to another indicative practical issue: parasitic searches, whish are now used by some of the ongoing SETI projects, are natural modus operandi for the observational search for the Galactic macro-engineering. This, of course, means a great increase in efficiency of operation, as well as a decrease in cost, especially when coupled with widely distributed processing, along the lines of ingenious SETI@home. However, this makes the role of creating solid theoretical groundwork for such projects much more delicate and important.

**Instead of Conclusions**

The Dysonian approach to search for other intelligent societies can be briefly summarized as follows. Even if they are not actively communicating with us, that does not imply that we cannot detect them and their astro-engineering activities. Their detection signatures may be much older than their communication signatures. Unless ATCs have taken great lengths to hide or disguise their IR detection signatures, the terrestrial observers should still be able to observe them at those wavelengths and those should be distinguishable from normal stellar spectra. (Ironically enough, surveys in the infrared have been proposed by one of the pioneers of microwave astronomy, Nobel-prize winner Charles H. Townes, although on somewhat different grounds; see Townes, 1983.) The same applies to other un-natural effects, like the antimatter-burning signatures (Harris, 1986, 2002), or recognizable transits of artificial objects (Arnold, 2005). Search for mega-projects such as Dyson shells, Jupiter Brains or stellar engines are most likely to be successful in the entire spectrum of SETI activities.

Some of the major differences between the orthodox and the Dysonian approach to SETI are laconically summarized in Table 13-1.

Table **Error! No text of specified style in document.**-1. A comparison between the orthodox and the Dysonian approach to SETI.

|  | orthodox SETI | Dysonian SETI |
|---|---|---|
| main object of search | intentional messages | artifacts and traces |
| working ATC model | biological | postbiological |
| window of opportunity | narrow | wide |
| prejudicates alien behavior | yes | no |
| communication | yes | no |
| interstellar travel | irrelevant | relevant |
| main working | radio (cm) | infrared |

|                        | orthodox SETI | Dysonian SETI |
|------------------------|---------------|---------------|
| frequencies            |               |               |
| natural mode of work   | active        | parasitic     |
| extragalactic SETI     | no            | yes           |

It would not be too pretentious to claim that the comparison of the two approaches favors the Dysonian approach. However, this approach has yet to achieve its legitimacy in the circles of SETI researchers. Bold and unconventional studies, such as Freitas', Harris', Arnold's, Slysh's, or the survey of Jugaku et al., represent still a small minority of the overall SETI research. We dare suggest that there is no real scientific reason for such situation: instead, it occurs due to excessive conservativeness, inertia of thought, overawe of the "founding fathers", or some combination of the three. Another, albeit extra-scientific, argument sometimes put forward in informal situations is that the massive pseudoscientific fringe surrounding SETI ("flying saucers" enthusiasts, archaeo-astronauts, and the like) would feel encouraged by relaxing the conservative tenets of the orthodox SETI. This argument is hard to evaluate due to its essentially social and extra-scientific nature. In any case, it gives far too much weight and influence to lunatics and pseudo-scientists than is tolerable in any serious scientific discipline. The unconventional approach with emphasis on search for ATCs' manifestations would lose nothing of the advantages of conventional SETI before detection (Tough, 1998), but the gains could be enormous.

Since it is the success in search we are after, it goes without saying that this assessment has nothing to do with SETI skeptics such as Tipler, Mayr, Carter, Conway Morris and others. Insofar as some of their arguments cogently contribute to our astrobiological understanding, they are indeed welcome, but the overall interpretation along the lines of "we are alone in the Galaxy" is a dangerous anthropocentric pretension. That the early SETI optimism was unjustified has nothing to do with serious and realistic work which is being done and will, it is to be hoped, continue to be done in the field. The Dysonian approach should not be construed as some nebulous "search for miracles", albeit cosmic miracles. Instead, it can be regarded as operationalization of the old epistemological dictum of Heraclitus of Ephesos: If you do not expect the unexpected, you will not find it; for it is hard to be sought out and difficult (fragment B18). Doesn't history of science teach us that such was the attitude of great innovators, revolutionaries, and original thinkers in general?

What is the benefit for macroengineers arising from the proposed cooperation with astrobiologists? First and foremost, they would be offered another fresh outlet for the "frustration" Dyson wrote about. In this multidisciplinary enterprise, fruitful and liberal exchange of ideas between very different specialists can be only beneficial for all. Different types of macro-engineering projects will have different astrobiological impact, and will require different detection methods and procedures, thus opening a wide and still almost entirely empty field for theoretical studies and modeling. The same issue has another side: a civilization wishing for some reason to avoid detection will likely refrain from at least some macro-engineering projects; whether this can apply to future humanity/posthumanity is for future social sciences and decision-making processes to determine. But in order to do so, detailed studies of macroprojects detectability need to be performed, exactly of the same kind required by astrobiology. Finally, the very limits of the concept of macro-engineering itself, vis-à-vis materials, energy, time and sophistication constraints are possible to probe only through the Dysonian approach to SETI; this is similar to the ways an astrophysicist learns more about our Sun, its structure, evolution and final destiny by observing billions of other stars in the Milky Way.

Overall, the greatest beneficiary may be the long-term future of intelligence itself. As Tsiolkovsky famously wrote in a 1911 letter: "The Earth is the cradle of the mind, but we cannot live forever in a cradle." Neither can others.